# E' paramagnetic centers in nanoporous synthetic opals – a probe for near surface protons

A. Galukhin, Y. Osin, A. Rodionov, G. Mamin, M. Gafurov, S. Orlinskii

Kazan Federal University, Kremlevskaya, 18, Kazan 420008, Russia

*E-mail: sergei.orlinskii@kpfu.ru*

The studies of chemical processes in spatially confined conditions are of interest from the fundamental and industrial points of view. By means of the W - band EPR and $^1$H Mims electron - nuclear double resonance (ENDOR) we show that the radiation-induced paramagnetic centers (E') in the synthetic nanoporous silica opals could be used as sensitive probes to investigate the surface modification and, potentially, reactions of polymerization in the confined by opal pores.

## 1. Introduction

The influence of nanoconfinement on physical processes, such as melting, crystallization and polymorphic transitions of both inorganic and organic substrates is well understood [1-4], while the studies of chemical processes in spatially confined conditions, especially in the condensed phase (solids and liquids), are mostly phenomenological in nature. A common theory that would explain the differences in reactivity for confined substrates, and in particular the mechanisms of how nanoconfinement affects chemical reactions, is still lacking. For example, two existing interpretations for confined radical polymerizations that associate the differences in reactivity with changes in the mobility of macromolecules or with direct influence of porous media on radical recombination, are contradictory [5-7]. Alternatively, increased reactivity of cyanate esters confined in silica nanopores was explained by the nanopore induced ordering [8-10].

In our study we hypothesize that X-ray induced stable paramagnetic centers in silica based porous media can provide with an important information about the surface (or near surface) state. It can potentially shed light on the nanoconfinement-induced ordering of substrates inside the nanoporous media. In addition, possible mechanisms of this ordering can be understood by exploiting the thermal polymerization of monocyanate esters (MCE) as a model process. Silica opals were chosen as a model porous medium that can be prepared by well established methods allowing to control pores sizes and surface properties [11].

It is known that parameters of the paramagnetic centers in solid materials are sensitive to the particle dimensionality / size [12] and their surface modification [13]. Therefore, electron paramagnetic resonance (EPR) and electron - nuclear double resonance (ENDOR) can serve as assistive tools for investigation the chemical processes in pores. In this work, using our experience with EPR / ENDOR characterization of $SiO_2$ [14] and $Al_2O_3$ [15] aerogels, neutron irradiated quartz crystals [16], surface chemistry of alumina [17, 18], nanodiamonds [13], nanosized apatites [19, 20] and processes of oxidation of heavy oil in the pores of synthetic opal [21], we show that E' centers in nanoporous silica opals created by X - ray irradiation can serve as appropriate paramagnetic probes to sense the surface modification and, potentially, near - surface chemical reactions.

## 2. Materials and Methods

Ammonium hydroxide solution (25% of $NH_3$, GOST 24147-80), tetraethylorthosilicate (TEOS, > 99.9 %, Aldrich Inc.) were purchased and used without additional purification. Absolute ethanol was obtained by consecutive distillations of 96 % ethanol over CaO and $CaH_2$.

Silica spheres were prepared by the two-step controllable growth technique based on regrowth of silica seeds [22]. At the first step the seed dispersion was prepared by mixing the solution of 4.5 mL (20 mmol) of TEOS in 80 mL of absolute ethanol with 3.0 mL (40 mmol) of ammonium hydroxide solution, 34 mL (1.9 mole) of deionized water in 80 mL of absolute ethanol. The reaction was carried out at 60 °C with mixing for 24 hours. Obtained dispersion contained approx. 20 nm sized silica nanoparticles. At the second step, 3.75 mL (46 mmol) of ammonium hydroxide solution was added to 100 mL of obtained seeds dispersion and then two solutions (270 mL of 1.5 M solution of TEOS in ethanol and 270 mL of 1.2 M $NH_3$, 16 M $H_2O$ solution in absolute ethanol) were dosed by MasterFlex peristaltic pump (Cole - Parmer) during 5 hours. The reaction was carried out at 60 °C with stirring for 24 hours. Silica particles were isolated by centrifugation (10000 rpm) and washed consistently with ethanol, water-ethanol solutions (water to ethanol ratios were 1:3, 1:1, 3:1) and deionized water. The dried silica spheres were then calcined in a furnace at 600 °C for 12 hours. Desired temperature was achieved at a heating rate of 60 °C × hour$^{-1}$.

Silica colloidal crystals were prepared by modified vertical deposition method based on isothermal heating evaporation-induced self-assembly (IHEISA) method [23]. Silica particles were dispersed in ethanol by sonication and glass beaker containing the dispersion was placed in a homemade setup as described in ref. [23] at 79.8 ˚C. Obtained colloidal crystals were carefully removed from the beaker's walls and calcined at 800 ˚C for 12 hours, desired temperature was achieved at a heating rate of 300 °C per hour. We used the modification of the vertical deposition method based on isothermal heating evaporation-induced self-assembly as a more reproducible and less time - consuming technique allowing to produce well - ordered colloidal crystal films [23].

The particle size in colloidal crystals was estimated by measuring 100 individual particles using scanning electron microscopy (SEM). SEM measurements were carried out using field - emission high - resolution scanning electron microscope Merlin Carl Zeiss (Figure 1). Observation images of surface morphology were obtained at accelerating voltage of incident electrons of 15 kV and current of 300 pA. The average size of the silica spheres was 93 ± 6 nm.

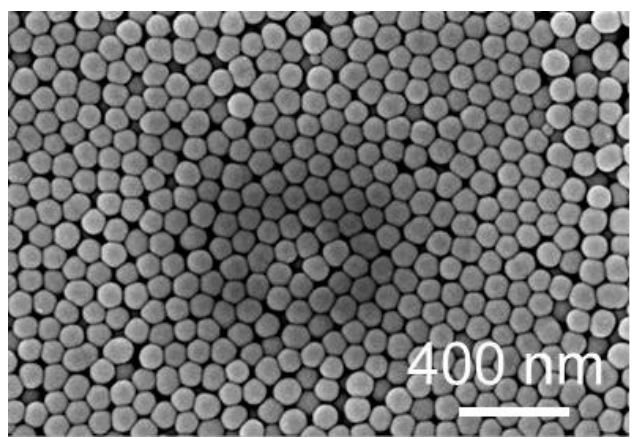

**Figure 1.** SEM image of synthesized opal sample.

Monocyanate ester with a structure shown in Figure 2 was used to study the interaction with opal surface. To create stable radicals in opal, X - ray irradiation of the URS - 55 source (U = 55 kV, I = 16 mA, W anticathode) at room temperature (RT) for 2 hours with the estimated dose of 15 kGy was exploited.

EPR investigations were done by using helium flow cryostats on the X - band ESP 300 (microwave frequency ν ≈ 9.5 GHz) and Bruker Elexsys W - band E680 (microwave frequency ν ≈ 93.5 GHz) spectrometers. We used conventional continuous wave (CW) and two-pulse field swept electron spin echo (FS ESE) measurements for the detection of EPR signal and the primary Hahn–echo (π/2 – τ – π – echo), where τ is the interpulse delay time of 240 ns, with initial π/2 and π pulse lengths of 32 and 64 ns, respectively. For the W - band ENDOR experiments we used special double (for nuclei and electron) cavities and Mims pulse sequence ($\pi/2 - \tau - \pi/2 - T - \pi/2$) with an additional radiofrequency (RF) pulse $\pi_{RF} = 18$ μs inserted between the second and third microwave $\pi/2$ pulses [24, 25]. RF frequency in our setup could be swept in the range of 1 - 200 MHz.

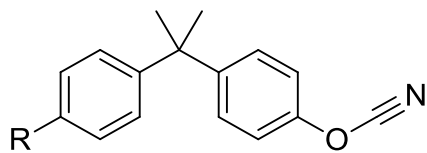

**Figure 2.** Structure of the monocyanate ester.

## 3. Results and Discussion

X-band EPR spectrum of opal sample at RT after irradiation (IOS – irradiated opal samples) is presented in Figure 3. No EPR signal was registered for the initial, non – irradiated samples within the sensitivity limits of our equipment.

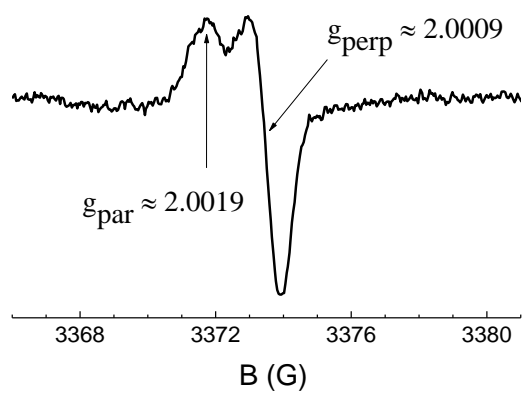

**Figure 2.** Room temperature X – band CW EPR of the X-ray irradiated opal film. Arrows point the g - factor components.

In natural and synthetic quartz the EPR signals are associated with oxygen vacancies (i.e., the so – called E defects that are commonly divided into E' and E" to denote the presence of one and two trapped electrons, respectively) known since the early 1950s [26, 27]. From Figure 3 it follows that the experimental spectrum is undoubtedly due to E' in disordered (powder) silica. In spite of decades of studies (mainly single crystal investigations), significant questions remain about the structural model for the E center in quartz and a – $SiO_2$. The exact EPR parameters of the paramagnetic E centers in the real species are hard to predict because of the diversity of silica crystal structures, various dimensions and surface state of the samples, presence of impurities, application of different types of the ionizing radiation (ultraviolet, neutron, gamma, X – ray), etc. Additionally, as it is stressed in one of the recent reviews on this topic, "structural models of radiation-induced defects obtained from single-crystal EPR analyses of crystalline $SiO_2$ (α – quartz) are often applicable to their respective analogues in amorphous silica (a – $SiO_2$), although significant differences are common" [26]. Therefore, no attempts of deep analysis of probable structural configurations for the detected paramagnetic centers are done.

Experimental X – band EPR spectra result from the paramagnetic center of axial symmetry with a pair of g – components $g_{par}$ = 2.0019(5) and $g_{perp}$ = 2.0009(5). However, no special procedures for their exact determination were done. These values are close to the usually reported for the neutron irradiated crystal quartz of $g_1$ = 2.0003, $g_2$ = 2.0006, $g_3$ = 2.0018 [16, 27] and for the $E'_2$ (I) center in the $^{60}$Co-irradiated α-quartz [28] but slightly higher, for example, than the values first derived in [29] for the germanium doped quartz after X-ray irradiation at RT ($g_1$ = 2.0011, $g_2$ = 1.9950, $g_3$ = 1.9939).

W – band spectra detected by FS ESE for the X – ray irradiated opal film filled with MCE and registered at T = 120 K are shown in Figure 4. As in the case of the X – band, the obtained signal could be described as a paramagnetic center of axial symmetry with $g_{par}$ = 2.0019(4) and $g_{perp}$ = 2.0009(3). No significant changes of the EPR signals were observed after saturation of the IOS by the proton containing liquid neither in X, nor in the W – band. Due to the higher spectral resolution, W – band experiments allow selectively saturate EPR transitions of the E' centers aligned parallel and perpendicular to the magnetic field for the ENDOR measurements.

$^1$H ENDOR spectra detected at T = 120 K for the water containing sample in the magnetic fields corresponding to the $g_{par}$ and $g_{perp}$ are presented in Figure 5. Well resolved ENDOR splittings were obtained for both values of the external magnetic field.

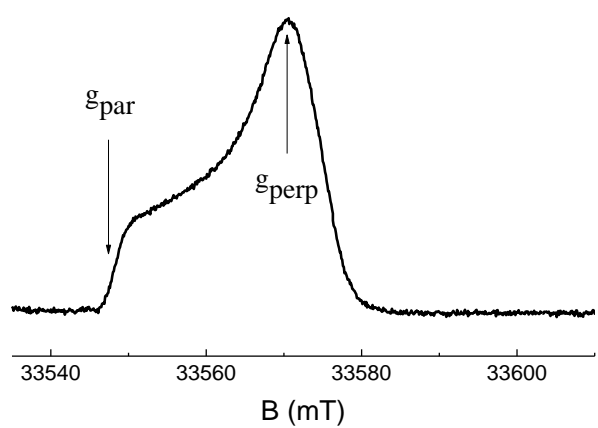

**Figure 4.** W – band FS ESE spectra of X – ray irradiated opal with the pores saturated by MCE detected at T = 120 K. Arrows point the positions of magnetic field corresponding to the g – factors components for which ENDOR spectra were measured.

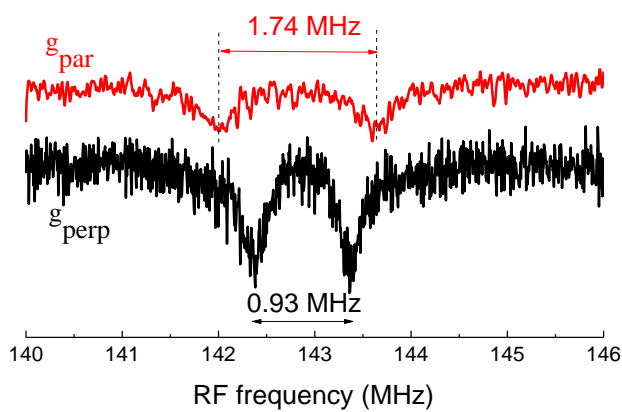

**Figure 5.** W – band $^1$H ENDOR spectra for the water filled opal sample at T = 120 K in the magnetic fields corresponding to the $g_{par}$ and $g_{pep}$. Values of ENDOR splittings are given.

The ENDOR patterns and ratio between the splittings for two orientations (roughly 2:1) allows to propose that the electron-proton hyperfine interaction originates mainly from the dipolar interaction. Indeed, in an assumption of the simple electron-proton point-dipole approximation which is applicable for the large distances between electron and nuclear spins, one can observe the splitting depending on the electron-proton distance *r*

$$|\Delta f| = g_S g_I \beta_S \beta_I (3\cos^2\Theta - 1)/r^3, \qquad (1)$$

where $g_S$, $g_I$, $\beta_S$, $\beta_I$ are the g-factors and Bohr's magnetons of the electron and nucleus, respectively, and $\Theta$ is the polar angle between the direction of the applied magnetic field and the line connecting the $S$ and $I$ spins [30]. The obtained value of $\Delta f$ corresponds to electron-proton distance of about 0.4 nm. Therefore, we can suggest that the registered in ENDOR measurements E – centres are located close to the material surface and "can sense" the protons at a particle surface or in its close vicinity. Additionally, opposite to the experimental results presented in paper [14], our experiments do not show the splittings of 3.7 mT (10.4 MHz) or 5.1 mT (14.3 MHz) neither in EPR nor in ENDOR spectra which were assigned to the hyperfine interactions with the residue protons in bulk silica (in $SiO_2$ lattice) that confirms our main conclusion.

## 4. Conclusion

By using the abilities of EPR / ENDOR spectroscopy, we have estimated the EPR parameters of the X – ray induced stable radicals in the nanoporous synthetic opals. We have shown that the stable radiation-induced paramagnetic centers can be exploited as sensitive surface probes. The obtained results can be used for studying the processes of chemical modifications of the opal surface in the micro- and nano-porous systems. It makes possible following the processes of chemical reactions in the confined media. The extracted EPR parameters could be also used for the elucidation of possible radiation-induced paramagnetic structures in the silica based materials.

## Acknowledgments

The work is financially supported by the Russian Science Foundation, Project # 18-13-00149.